\documentclass[aps,prd,reprint,superscriptaddress,showpacs,showkeys]{revtex4-1}

\usepackage{graphicx}
\usepackage{dcolumn}
\usepackage{bm}
\usepackage{natbib}
\usepackage{color}
\usepackage{subfigure}
\usepackage{array}
\usepackage{url}
\usepackage{hyperref}
\usepackage{multirow}

\begin{document}
\title{Performances of a prototype point-contact germanium detector immersed in liquid nitrogen for light dark matter search}

\affiliation{Key Laboratory of Particle and Radiation Imaging (Ministry of Education) and Department of Engineering Physics, Tsinghua University, Beijing 100084}
\affiliation{College of Nuclear Science and Technology, Beijing Normal University, Beijing 100875}
\affiliation{Institute of Physics, Academia Sinica, Taipei 11529}
\affiliation{Department of Physics, Dokuz Eyl\"{u}l University, \.{I}zmir 35160}
\affiliation{Department of Physics, Tsinghua University, Beijing 100084}
\affiliation{NUCTECH Company, Beijing 100084}
\affiliation{YaLong River Hydropower Development Company, Chengdu 610051}
\affiliation{College of Physical Science and Technology, Sichuan University, Chengdu 610064}
\affiliation{Department of Nuclear Physics, China Institute of Atomic Energy, Beijing 102413}
\affiliation{School of Physics, Nankai University, Tianjin 300071}
\affiliation{Department of Physics, Banaras Hindu University, Varanasi 221005}
\affiliation{Department of Physics, Beijing Normal University, Beijing 100875}
\author{H. Jiang}
\affiliation{Key Laboratory of Particle and Radiation Imaging (Ministry of Education) and Department of Engineering Physics, Tsinghua University, Beijing 100084}
\author{L.~T.~Yang}\email{Corresponding author: yanglt@mail.tsinghua.edu.cn}
\affiliation{Key Laboratory of Particle and Radiation Imaging (Ministry of Education) and Department of Engineering Physics, Tsinghua University, Beijing 100084}
\author{Q. Yue}\email{Corresponding author: yueq@mail.tsinghua.edu.cn}
\affiliation{Key Laboratory of Particle and Radiation Imaging (Ministry of Education) and Department of Engineering Physics, Tsinghua University, Beijing 100084}
\author{K.~J.~Kang}
\affiliation{Key Laboratory of Particle and Radiation Imaging (Ministry of Education) and Department of Engineering Physics, Tsinghua University, Beijing 100084}
\author{J.~P.~Cheng}
\affiliation{Key Laboratory of Particle and Radiation Imaging (Ministry of Education) and Department of Engineering Physics, Tsinghua University, Beijing 100084}
\affiliation{College of Nuclear Science and Technology, Beijing Normal University, Beijing 100875}
\author{Y.~J.~Li}
\affiliation{Key Laboratory of Particle and Radiation Imaging (Ministry of Education) and Department of Engineering Physics, Tsinghua University, Beijing 100084}
\author{H.~T.~Wong}
\altaffiliation{Participating as a member of TEXONO Collaboration}
\affiliation{Institute of Physics, Academia Sinica, Taipei 11529}
\author{M. Agartioglu}
\altaffiliation{Participating as a member of TEXONO Collaboration}
\affiliation{Institute of Physics, Academia Sinica, Taipei 11529}
\affiliation{Department of Physics, Dokuz Eyl\"{u}l University, \.{I}zmir 35160}
\author{H.~P.~An}
\affiliation{Key Laboratory of Particle and Radiation Imaging (Ministry of Education) and Department of Engineering Physics, Tsinghua University, Beijing 100084}
\affiliation{Department of Physics, Tsinghua University, Beijing 100084}
\author{J.~P.~Chang}
\affiliation{NUCTECH Company, Beijing 100084}
\author{J.~H.~Chen}
\altaffiliation{Participating as a member of TEXONO Collaboration}
\affiliation{Institute of Physics, Academia Sinica, Taipei 11529}
\author{Y.~H.~Chen}
\affiliation{YaLong River Hydropower Development Company, Chengdu 610051}
\author{Z.~Deng}
\affiliation{Key Laboratory of Particle and Radiation Imaging (Ministry of Education) and Department of Engineering Physics, Tsinghua University, Beijing 100084}
\author{Q.~Du}
\affiliation{College of Physical Science and Technology, Sichuan University, Chengdu 610064}
\author{H.~Gong}
\affiliation{Key Laboratory of Particle and Radiation Imaging (Ministry of Education) and Department of Engineering Physics, Tsinghua University, Beijing 100084}
\author{L. He}
\affiliation{NUCTECH Company, Beijing 100084}
\author{J.~W.~Hu}
\affiliation{Key Laboratory of Particle and Radiation Imaging (Ministry of Education) and Department of Engineering Physics, Tsinghua University, Beijing 100084}
\author{Q.~D.~Hu}
\affiliation{Key Laboratory of Particle and Radiation Imaging (Ministry of Education) and Department of Engineering Physics, Tsinghua University, Beijing 100084}
\author{H.~X.~Huang}
\affiliation{Department of Nuclear Physics, China Institute of Atomic Energy, Beijing 102413}
\author{L.~P.~Jia}
\affiliation{Key Laboratory of Particle and Radiation Imaging (Ministry of Education) and Department of Engineering Physics, Tsinghua University, Beijing 100084}
\author{H.~B.~Li}
\altaffiliation{Participating as a member of TEXONO Collaboration}
\affiliation{Institute of Physics, Academia Sinica, Taipei 11529}
\author{H. Li}
\affiliation{NUCTECH Company, Beijing 100084}
\author{J.~M.~Li}
\affiliation{Key Laboratory of Particle and Radiation Imaging (Ministry of Education) and Department of Engineering Physics, Tsinghua University, Beijing 100084}
\author{J.~Li}
\affiliation{Key Laboratory of Particle and Radiation Imaging (Ministry of Education) and Department of Engineering Physics, Tsinghua University, Beijing 100084}
\author{X.~Li}
\affiliation{Department of Nuclear Physics, China Institute of Atomic Energy, Beijing 102413}
\author{X.~Q.~Li}
\affiliation{School of Physics, Nankai University, Tianjin 300071}
\author{Y.~L.~Li}
\affiliation{Key Laboratory of Particle and Radiation Imaging (Ministry of Education) and Department of Engineering Physics, Tsinghua University, Beijing 100084}

\author {B. Liao}
\affiliation{College of Nuclear Science and Technology, Beijing Normal University, Beijing 100875}

\author{F.~K.~Lin}
\altaffiliation{Participating as a member of TEXONO Collaboration}
\affiliation{Institute of Physics, Academia Sinica, Taipei 11529}
\author{S.~T.~Lin}
\affiliation{College of Physical Science and Technology, Sichuan University, Chengdu 610064}
\author{S.~K.~Liu}
\affiliation{College of Physical Science and Technology, Sichuan University, Chengdu 610064}

\author {Y.~D.~Liu}
\affiliation{College of Nuclear Science and Technology, Beijing Normal University, Beijing 100875}
\author {Y.~Y.~Liu}
\affiliation{College of Nuclear Science and Technology, Beijing Normal University, Beijing 100875}

\author{Z.~Z.~Liu}
\affiliation{Key Laboratory of Particle and Radiation Imaging (Ministry of Education) and Department of Engineering Physics, Tsinghua University, Beijing 100084}
\author{H.~Ma}\email{Corresponding author: mahao@mail.tsinghua.edu.cn}
\affiliation{Key Laboratory of Particle and Radiation Imaging (Ministry of Education) and Department of Engineering Physics, Tsinghua University, Beijing 100084}
\author{J.~L.~Ma}
\affiliation{Key Laboratory of Particle and Radiation Imaging (Ministry of Education) and Department of Engineering Physics, Tsinghua University, Beijing 100084}
\affiliation{Department of Physics, Tsinghua University, Beijing 100084}
\author{H.~Pan}
\affiliation{NUCTECH Company, Beijing 100084}
\author{J.~Ren}
\affiliation{Department of Nuclear Physics, China Institute of Atomic Energy, Beijing 102413}
\author{X.~C.~Ruan}
\affiliation{Department of Nuclear Physics, China Institute of Atomic Energy, Beijing 102413}
\author{B. Sevda}
\altaffiliation{Participating as a member of TEXONO Collaboration}
\affiliation{Institute of Physics, Academia Sinica, Taipei 11529}
\affiliation{Department of Physics, Dokuz Eyl\"{u}l University, \.{I}zmir 35160}
\author{V.~Sharma}
\altaffiliation{Participating as a member of TEXONO Collaboration}
\affiliation{Institute of Physics, Academia Sinica, Taipei 11529}
\affiliation{Department of Physics, Banaras Hindu University, Varanasi 221005}
\author{M.~B.~Shen}
\affiliation{YaLong River Hydropower Development Company, Chengdu 610051}
\author{L.~Singh}
\altaffiliation{Participating as a member of TEXONO Collaboration}
\affiliation{Institute of Physics, Academia Sinica, Taipei 11529}
\affiliation{Department of Physics, Banaras Hindu University, Varanasi 221005}
\author{M.~K.~Singh}
\altaffiliation{Participating as a member of TEXONO Collaboration}
\affiliation{Institute of Physics, Academia Sinica, Taipei 11529}
\affiliation{Department of Physics, Banaras Hindu University, Varanasi 221005}

\author {T.~X.~Sun}
\affiliation{College of Nuclear Science and Technology, Beijing Normal University, Beijing 100875}

\author{C.~J.~Tang}
\affiliation{College of Physical Science and Technology, Sichuan University, Chengdu 610064}
\author{W.~Y.~Tang}
\affiliation{Key Laboratory of Particle and Radiation Imaging (Ministry of Education) and Department of Engineering Physics, Tsinghua University, Beijing 100084}
\author{Y.~Tian}
\affiliation{Key Laboratory of Particle and Radiation Imaging (Ministry of Education) and Department of Engineering Physics, Tsinghua University, Beijing 100084}

\author {G.~F.~Wang}
\affiliation{College of Nuclear Science and Technology, Beijing Normal University, Beijing 100875}

\author{J.~M.~Wang}
\affiliation{YaLong River Hydropower Development Company, Chengdu 610051}
\author{L.~Wang}
\affiliation{Department of Physics, Beijing Normal University, Beijing 100875}
\author{Q.~Wang}
\affiliation{Key Laboratory of Particle and Radiation Imaging (Ministry of Education) and Department of Engineering Physics, Tsinghua University, Beijing 100084}
\affiliation{Department of Physics, Tsinghua University, Beijing 100084}
\author{Y.~Wang}
\affiliation{Key Laboratory of Particle and Radiation Imaging (Ministry of Education) and Department of Engineering Physics, Tsinghua University, Beijing 100084}
\affiliation{Department of Physics, Tsinghua University, Beijing 100084}
\author{S.~Y.~Wu}
\affiliation{YaLong River Hydropower Development Company, Chengdu 610051}
\author{Y.~C.~Wu}
\affiliation{Key Laboratory of Particle and Radiation Imaging (Ministry of Education) and Department of Engineering Physics, Tsinghua University, Beijing 100084}
\author{H.~Y.~Xing}
\affiliation{College of Physical Science and Technology, Sichuan University, Chengdu 610064}
\author{Y.~Xu}
\affiliation{School of Physics, Nankai University, Tianjin 300071}
\author{T.~Xue}
\affiliation{Key Laboratory of Particle and Radiation Imaging (Ministry of Education) and Department of Engineering Physics, Tsinghua University, Beijing 100084}
\author{S.~W.~Yang}
\altaffiliation{Participating as a member of TEXONO Collaboration}
\affiliation{Institute of Physics, Academia Sinica, Taipei 11529}
\author{N.~Yi}
\affiliation{Key Laboratory of Particle and Radiation Imaging (Ministry of Education) and Department of Engineering Physics, Tsinghua University, Beijing 100084}
\author{C.~X.~Yu}
\affiliation{School of Physics, Nankai University, Tianjin 300071}
\author{H.~J.~Yu}
\affiliation{NUCTECH Company, Beijing 100084}
\author{J.~F.~Yue}
\affiliation{YaLong River Hydropower Development Company, Chengdu 610051}
\author{X.~H.~Zeng}
\affiliation{YaLong River Hydropower Development Company, Chengdu 610051}
\author{M.~Zeng}
\affiliation{Key Laboratory of Particle and Radiation Imaging (Ministry of Education) and Department of Engineering Physics, Tsinghua University, Beijing 100084}
\author{Z.~Zeng}
\affiliation{Key Laboratory of Particle and Radiation Imaging (Ministry of Education) and Department of Engineering Physics, Tsinghua University, Beijing 100084}

\author {F.~S.~Zhang}
\affiliation{College of Nuclear Science and Technology, Beijing Normal University, Beijing 100875}

\author{Y.~H.~Zhang}
\affiliation{YaLong River Hydropower Development Company, Chengdu 610051}
\author{M.~G.~Zhao}
\affiliation{School of Physics, Nankai University, Tianjin 300071}
\author{J.~F.~Zhou}
\affiliation{YaLong River Hydropower Development Company, Chengdu 610051}
\author{Z.~Y.~Zhou}
\affiliation{Department of Nuclear Physics, China Institute of Atomic Energy, Beijing 102413}
\author{J.~J.~Zhu}
\affiliation{College of Physical Science and Technology, Sichuan University, Chengdu 610064}
\author{Z.~H.~Zhu}
\affiliation{YaLong River Hydropower Development Company, Chengdu 610051}

\collaboration{CDEX Collaboration}
\noaffiliation

\date{August 30, 2018}

\begin{abstract}
The CDEX-10 experiment searches for light weakly-interacting massive particles, a form of dark matter, at the China JinPing underground laboratory, where approximately 10 kg of germanium detectors are arranged in an array and immersed in liquid nitrogen. Herein, we report on the experimental apparatus, detector characterization, and spectrum analysis of one prototype detector. Owing to the higher rise-time resolution of the CDEX-10 prototype detector as compared with CDEX-1B, we identified the origin of an observed category of extremely fast events. For data analysis of the CDEX-10 prototype, we introduced and applied an improved bulk/surface event discrimination method. The results of the new method were compared to those of the CDEX-1B spectrum. Both sets of results showed good consistency in the 0--12 keVee energy range, except for the 8.0 keV K-shell X-ray peak from the external copper.
\end{abstract}

\maketitle


\section{Introduction}

Astronomical observations consistently indicate that nearly 85\% of the matter content of our universe is long-lived, non-luminous dark matter (DM)~\cite{PDG2017}. Further research on DM is essential to understand the origin of matter and the formation and evolution of the Universe. Weakly interacting massive particles (WIMPs, denoted as $\chi$) are among the most promising DM candidates. Light WIMPS ($m_{\chi} <10$ GeV/c$^2$) have been searched in various experiments using germanium, silicon, and CaWO$_{4}$ detectors~\cite{CoGeNT_PRL_2011,cdmslite, CDMS_Si_PRL_2013, Angloher2012}. However, no WIMPs have yet been found.

Established in 2009, the China Dark Matter Experiment (CDEX) Collaboration~\cite{CDEX_CPC_2013} searched for light WIMPS using $p$-type point-contact germanium ($p$PCGe) detectors at the China JinPing underground laboratory (CJPL)~\cite{cjpl}. The CJPL is the world’s deepest underground laboratory, with a cosmic ray muon flux of 61.7 yr$^{-1}$m$^{-2}$~\cite{WUYC_CPC_2013}. Owing to the low physics analysis threshold (160 eVee, where``eVee" represents electron equivalent energy), the excellent energy resolution and low background characteristics of the $p$PCGe detectors, the CDEX collaboration has achieved a series of competitive results on light WIMP searches~\cite{cdex1,cdex12014,cdex12016,cdex12018}. The CDEX-1 experiment, which uses 1 kg-scale single-element $p$PCGe detectors, has proceeded in two stages: CDEX-1A and CDEX-1B.

In CDEX-1A and CDEX-1B, the germanium detector was cooled by a cold finger connected to a 30-L dewar filled with liquid nitrogen (LN$_{2}$). However, upscaling the mass of the germanium detector system, to further improve the sensitivity, is not easily accomplished. Toward a future ton-scale WIMP search experiment, the second-generation CDEX experiment (CDEX-10) has been established with a total detector crystal mass of approximately 10 kg. The CDEX-10 experiment has three encapsulated detector strings called C10A, C10B, and C10C, each comprising three point-contact germanium detectors encapsulated in a copper vacuum tube and immersed directly into LN$_{2}$. The high-Z shielding of the germanium crystals can be replaced by a low-Z material such as LN$_{2}$ or liquid argon, thereby providing an integrated shielding and cooling system that better controls the radioactive background~\cite{gerda,heusser}. The CDEX-10 experiment represents an important step from the single element detector with a cold finger to a detector array immersed in LN$_{2}$. It focuses on arraying technologies, the LN$_{2}$ cooling system, and a background understanding of the prototype $p$PCGe detectors developed via the CDEX-1 technique. The present article reports on the experimental apparatus, detector characterization, and spectrum analysis of a prototype germanium detector of the CDEX-10 system.

\section{Experimental Setup}

The CDEX-10 system resides in a polyethylene room with 1-m-thick walls at the CJPL. The whole system comprises a stainless steel LN$_{2}$ tank with integrated copper shielding, the 10 kg point-contact germanium detector array, and a data acquisition (DAQ) system. 

\begin{figure}[!htbp]
\centering\includegraphics[width=\columnwidth]{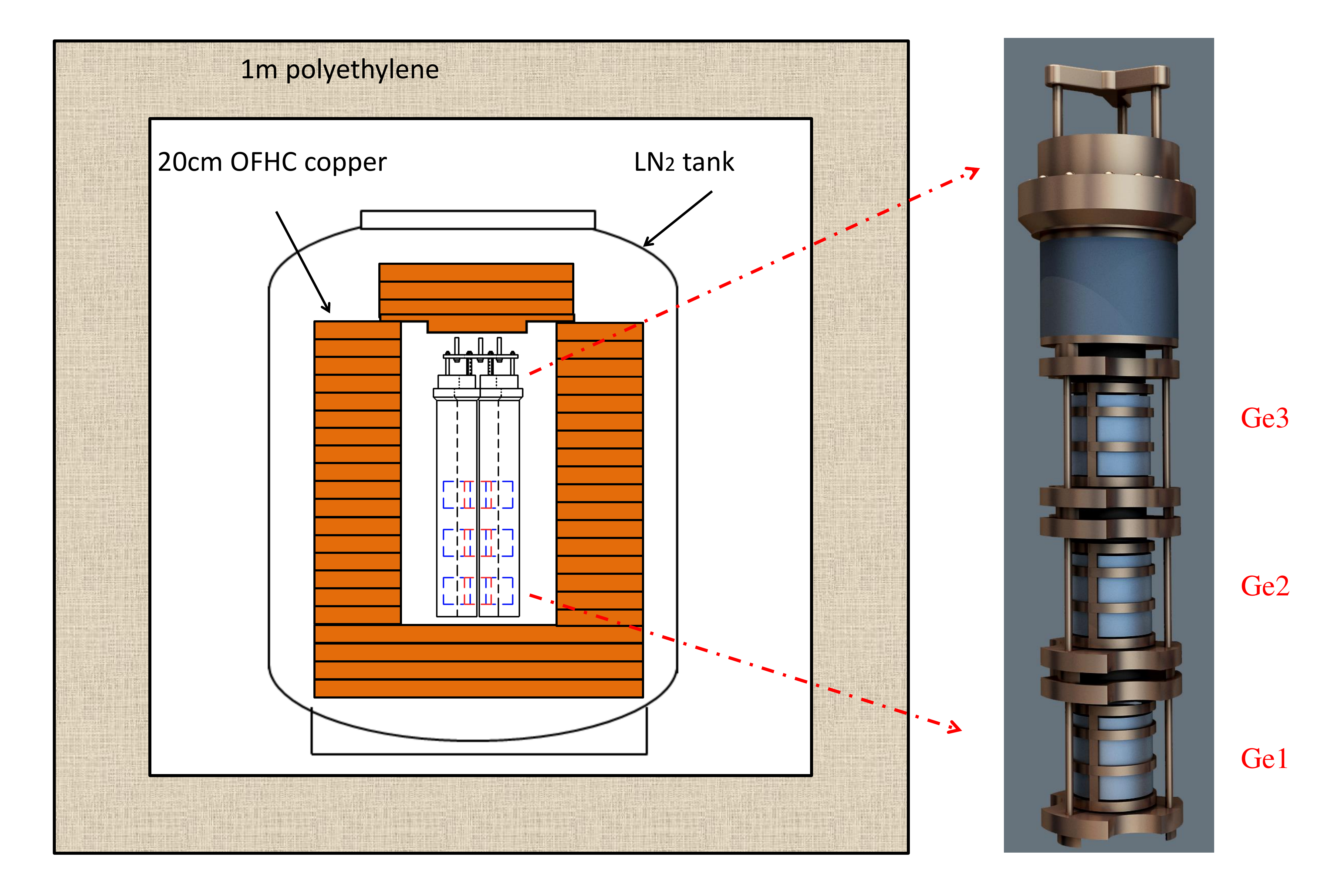}
\caption{Configuration of the CDEX-10 experimental setup (left) and detector string layout (right).}
\label{fig::detector}
\end{figure}

The experiment is operated at CJPL in a stainless-steel LN$_{2}$ tank with an outer diameter of 1.5 m and a height of 1.9 m (see Figure~\ref{fig::detector}). From outside to inside, the passive shielding system comprises 2400 m of rock overburden, a 1-m-thick polyethylene layer, and 20-cm-thick high-purity oxygen-free high-conductivity (OFHC) copper immersed in the LN$_{2}$ surrounding the detector array. In this configuration, the germanium detectors are directly cooled to a stable operating temperature by the LN$_{2}$. 

Each detector string in CDEX-10 contains 3 point-contact germanium detectors denoted as Ge1--Ge3 from bottom to top, as shown in Figure~\ref{fig::detector}. Each germanium crystal is approximately 62 mm in diameter and 62 mm high. 

\begin{figure*}[!htbp]
\centering\includegraphics[width=1.7\columnwidth]{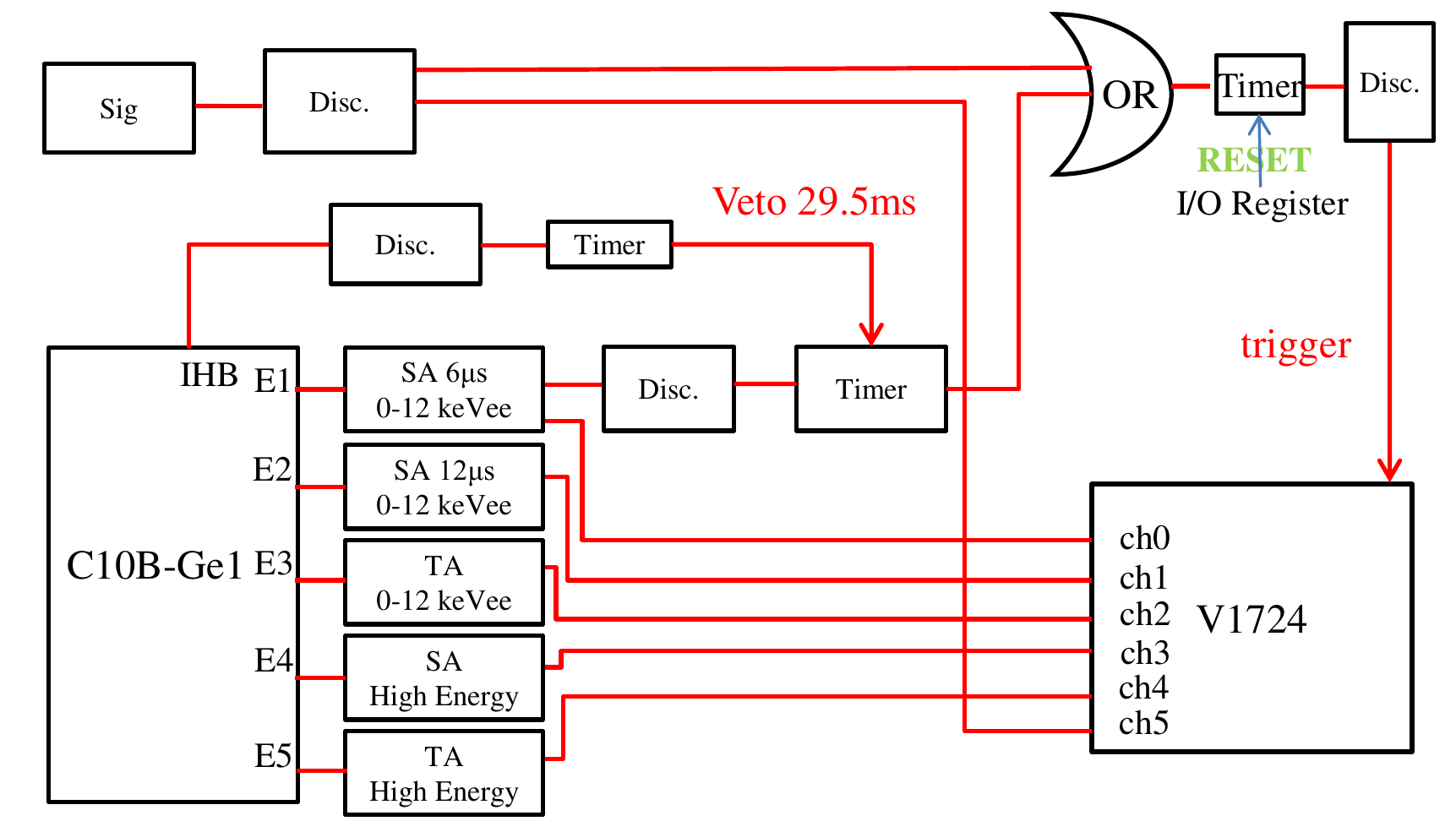}
\caption{Schematic of the DAQ system of the C10B-Ge1 detector. Sig: signal generator, Disc: discriminator, SA: shaping amplifier, TA: timing amplifier.}
\label{fig::daq}
\end{figure*}

Figure~\ref{fig::daq} is a schematic of the DAQ system of the C10B-Ge1 detector. The signals from the $p^{+}$ point-contact electrode of C10B-Ge1 are fed into a pulsed reset preamplifier, which exports five identical output signals. Each of the three shaping amplifiers and two timing amplifiers receive one of these outputs for further processing and digitization. The two high-gain shaping amplifiers $S_{p6}$ and $S_{p12}$ cover the 0--12 keVee energy range for light DM analysis with shaping times of 6 $\mu$s and 12 $\mu$s, respectively, and Gaussian filtering. The high-gain timing amplifier ($T_{p}$) measures the rise-time of signals in the 0--12 keVee energy range to distinguish between bulk and surface events. Finally, the low-gain shaping amplifier and low-gain timing amplifier cover the high-energy range for background understanding. The system is triggered by the $S_{p6}$ signal loaded onto a leading-edge discriminator. In addition, 0.05-Hz random trigger (RT) signals are generated by a signal generator. Using the RT signals, we can determine the zero point energy resolution, estimate the dead time of the DAQ system, and cut the signals that are uncorrelated to any energy deposit. The output signals of the above amplifiers were digitized via 14-bit 100-MHz flash analog-to-digital converters (FADCs). Once triggered by a $S_{p6}$ or an RT signal, the FADC records the raw output pulses from those amplifiers, with a time window of 120 $\mu$s. Notably, the pulsed reset preamplifier generates discharge signals (Inhibit)~\cite{WangL2017}, which can induce significant noises and false triggers in the subsequent time period. To avoid this problem, each Inhibit signal was followed by a 29.5-ms veto signal supplied by a timer.

\section{Detector characterization}

\begin{table*}[!htbp]
\begin{ruledtabular}
\caption{Selected parameters of the C10B-Ge1, CDEX-1A, and CDEX-1B detectors.}
\label{tab:comparison}
\centering
\begin{tabular}{lccc}
Detector & CDEX-1A & CDEX-1B & C10B-Ge1\\
\hline
Background level (cpkkd$^{ a)}$) @ 2 keVee  & 3.5 & 3.0 & 2.5\\
RT sigma (eVee) & 55 & 31 & 32 \\
Physics analysis threshold (eVee) & 475 & 160 & 160 \\
Combined signal efficiency at threshold & 80\% & 17\% & 4.5\%\\
FWHM$^{ b)}$ of 10.37 keV peak (eVee)  & 207 & 177 & 219\\
\end{tabular}
\end{ruledtabular}
\leftline{~$^{ a)}$counts per kg per keVee per day (cpkkd);~$^{ b)}$Full Width at Half Maximum (FWHM).}
\end{table*}

\begin{figure}[!htbp]
\centering\includegraphics[width=0.95\columnwidth]{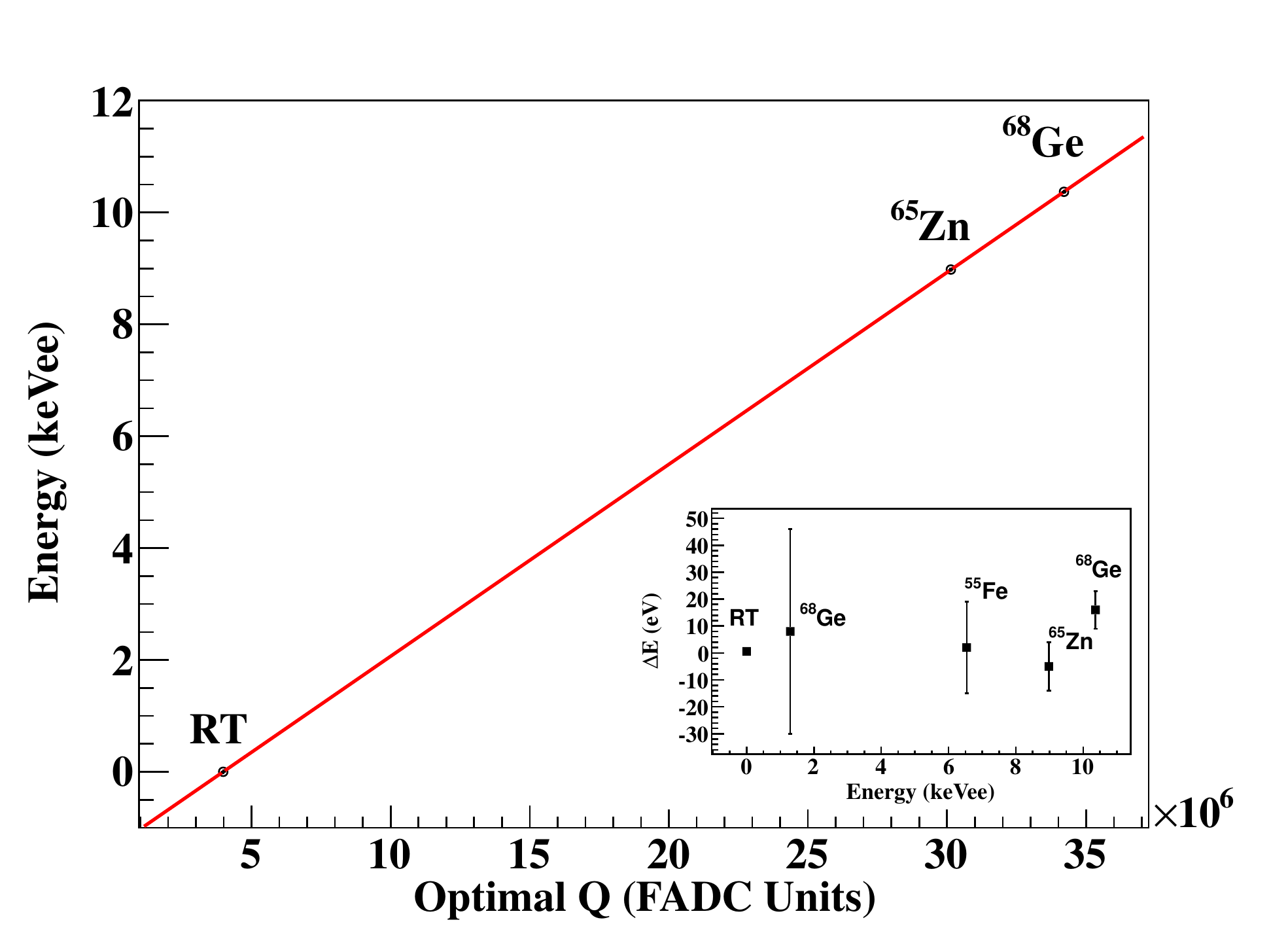}
\caption{Energy calibration of the C10B-Ge1 detector in the low-energy region using the 10.37 keV peak of $^{68}$Ge and the 8.98 keV peak of $^{65}$Zn, and the zero energy defined by the RT events. The inset shows the energy deviations of these calibrated peaks, the 1.30 keV peak of $^{68}$Ge, and the 6.54 keV peak of $^{55}$Fe. The energy deviation from linearity is smaller than 16 eVee.}
\label{fig::energycali_low}
\end{figure}

In the first physics operation of CDEX-10, the data were collected by C10B-Ge1 owing to its lower noise (lower RT sigma) and background level than that of the other detectors (see Table~\ref{tab:comparison}). 

The energy calibration is based on the parameters extracted from the digitized pulses. The energy of the C10B-Ge1 detector is defined in terms of the optimal integrated area of the pulses from $S_{p12}$, which are strongly proportional to energy in the low-energy region. 

As the C10B-Ge1 detector was placed at CJPL for over 2 years prior to commencing the formal data acquisition, most of the internal cosmogenic X-ray peaks had poor statistics~\cite{MaJL2019}. For energy calibration in the 0--12 keVee energy range, we used the main internal cosmogenic X-ray peaks (the 10.37 keV peak of $^{68}$Ge and the 8.98 keV peak of $^{65}$Zn), and the zero energy defined by the RT events (see Figure~\ref{fig::energycali_low}). The energy linearity was checked by two peaks: the 1.30 keV peak from $^{68}$Ge and the 6.54 keV peak from $^{55}$Fe. The energy deviation from the linear fit was less than 16 eVee.

C10B-Ge1 achieved an energy threshold of 160 eVee with a combined signal efficiency of 4.5\%. The energy thresholds of CDEX-1A and CDEX-1B are also shown in Table~\ref{tab:comparison}. 

\begin{figure}[!htbp]
\centering\includegraphics[width=\columnwidth]{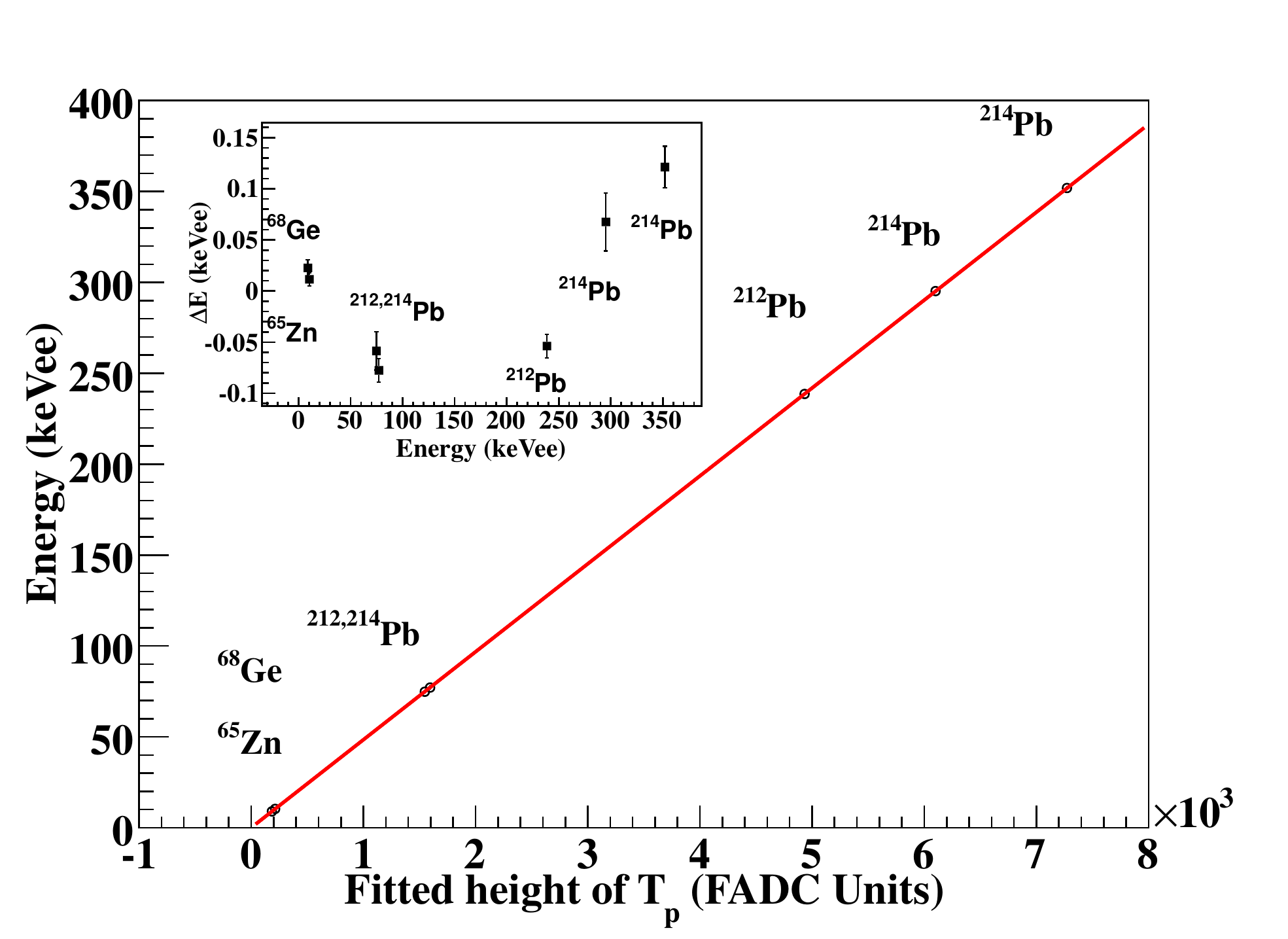}
\caption{Energy calibration of the C10B-Ge1 detector in the high-energy region. The inset shows the energy deviation of the calibrated peak. The energy deviations are below 0.15 keVee in the 0--350 keVee energy range.}
\label{fig::energycali_high}
\end{figure}

In the high-energy region, the spectrum containing several K-shell X-rays (KX) and $\gamma$-rays from $^{65}$Zn, $^{68}$Ge, and $^{212,~214}$Pb was calibrated using the fitted height of the $T_{p}$ pulse. The calibration is plotted in Figure~\ref{fig::energycali_high}. The energy deviation from linearity was below 0.15 keVee in the energy range 0--350 keVee.

\begin{figure}[!htbp]
\centering\includegraphics[width=\columnwidth]{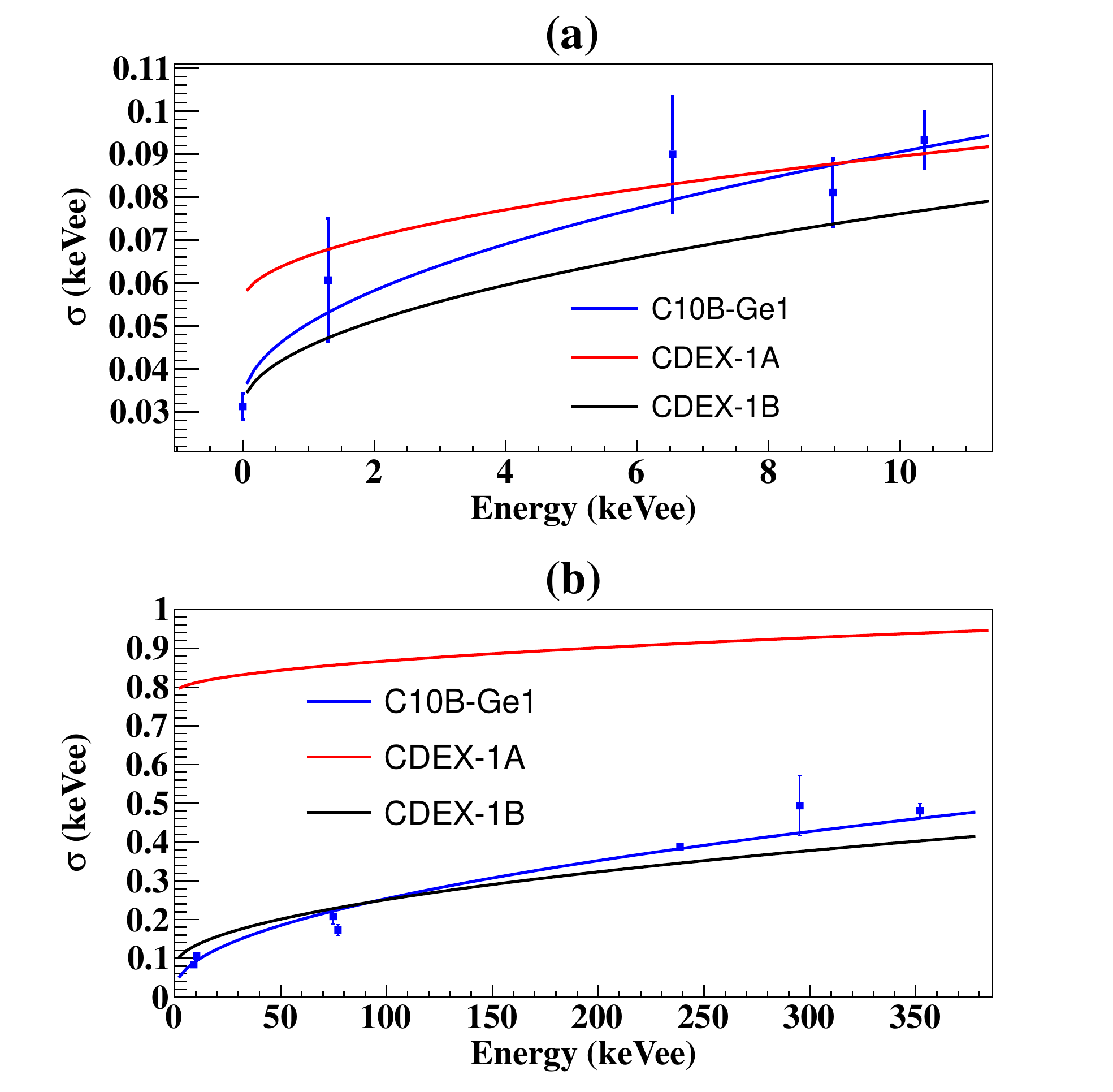}
\caption{Energy resolutions of C10B-Ge1 in the low- and high-energy ranges. The resolutions of CDEX-1A and CDEX-1B are also shown for comparison.}
\label{fig::sigma_low_high}
\end{figure}

The energy resolutions of C10B-Ge1, CDEX-1A, and CDEX-1B are compared in Figure~\ref{fig::sigma_low_high} and Table~\ref{tab:comparison}. In the lower-energy region (0--2 keVee), which is relevant to light DM search experiments, the energy resolutions of the C10B-Ge1 and CDEX-1B detectors are comparable.

\newpage

\section{Spectrum Analysis}
\begin{figure}[!htbp]
\centering\includegraphics[width=\columnwidth]{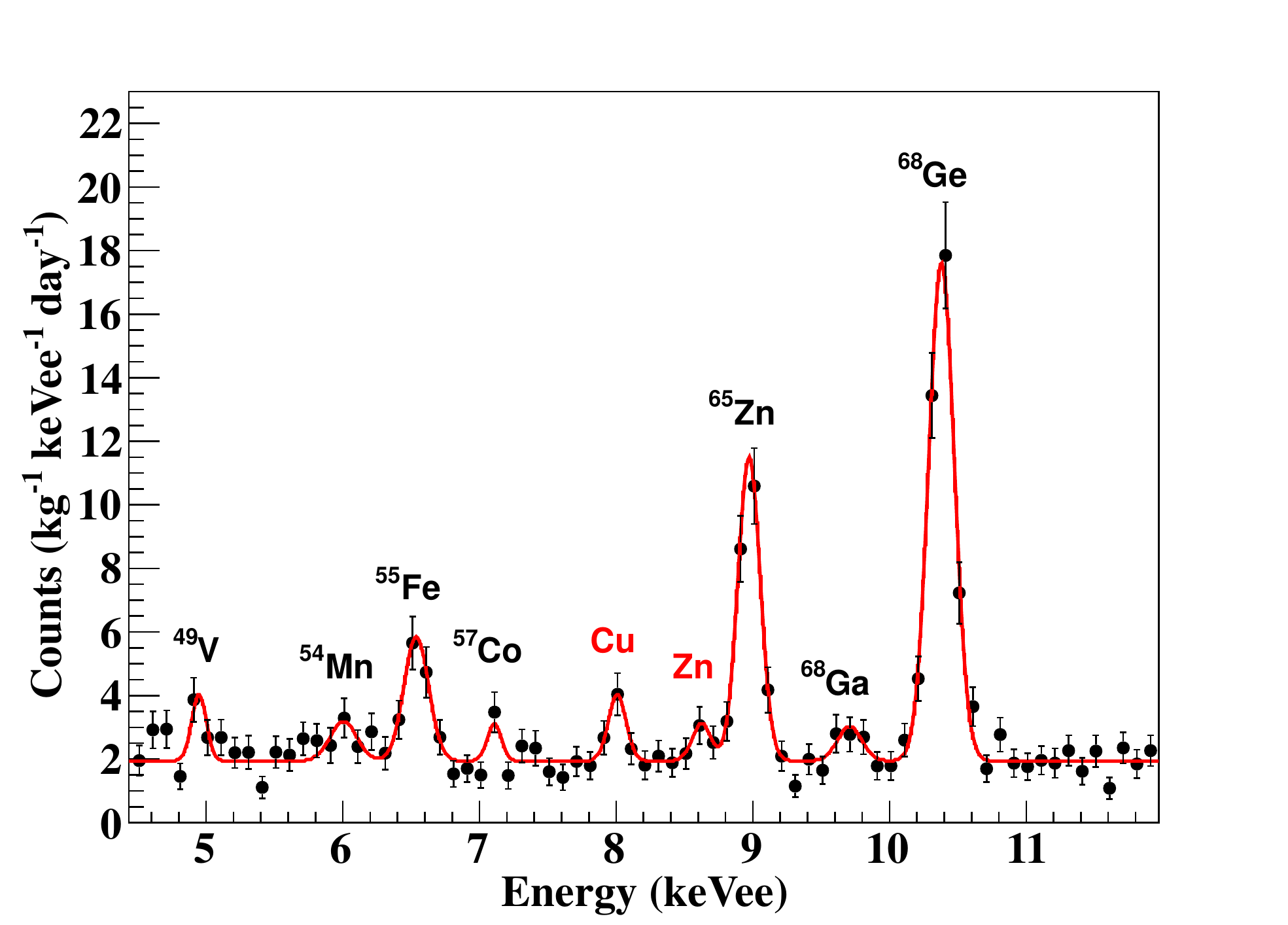}
\caption{Spectrum in the 4.5--12.0 keVee energy range. The red curve is the fitting result obtained using Eq.~(\ref{eq:KXfit}).}
\label{fig::spectrum_low}
\end{figure}

Figure~\ref{fig::spectrum_low} depicts the background spectrum of the C10B-Ge1 detector at low energies. Several KX peaks from cosmogenic isotopes were observed and identified between 4.5--12.0 keVee.

The intensities of the KX peaks were obtained by globally fitting the spectrum from 4.5 keVee to 12.0 keVee, using the sum of a flat background assumption and nine Gaussian functions:
\begin{eqnarray}
f(E) = p_0+\sum_{i}{a_i\times\frac{1}{\sqrt{2\pi}\sigma_i}exp\left(-\frac{(E-E_i)^2}{2\sigma_i^2}\right)},
\label{eq:KXfit}
\end{eqnarray}
where $p_{0}$ denotes the flat background, $i$ indexes the contributions from the internal isotopes $^{68}$Ge, $^{68}$Ga, $^{65}$Zn, $^{57}$Co, $^{55}$Fe, $^{54}$Mn and $^{49}$V, and Zn and Cu denote the contributions from external x-rays. $a_{i}$, $\sigma_{i}$ and $E_{i}$ represent the intensity, energy resolution, and calibrated energy, respectively, of an individual peak.

In the 4.5--12.0 keVee range, the fitted background level p$_0$ was 1.9 cpkkd. For comparison, the background level of CDEX-1A was obtained as 2.8 cpkkd, whereas CDEX-1B exhibited a slight rise ~\cite{cdex12018}.

The seven KX peaks were contributed by the internal cosmogenic isotopes $^{68}$Ge, $^{68}$Ga, $^{65}$Zn, $^{57}$Co, $^{55}$Fe, $^{54}$Mn, $^{49}$V in the germanium crystal. The remaining two were X-rays from the Cu and Zn materials close to the crystal, which are excited by high-energy $\gamma$-rays. 

\begin{figure}[!htbp]
\centering\includegraphics[width=0.95\columnwidth]{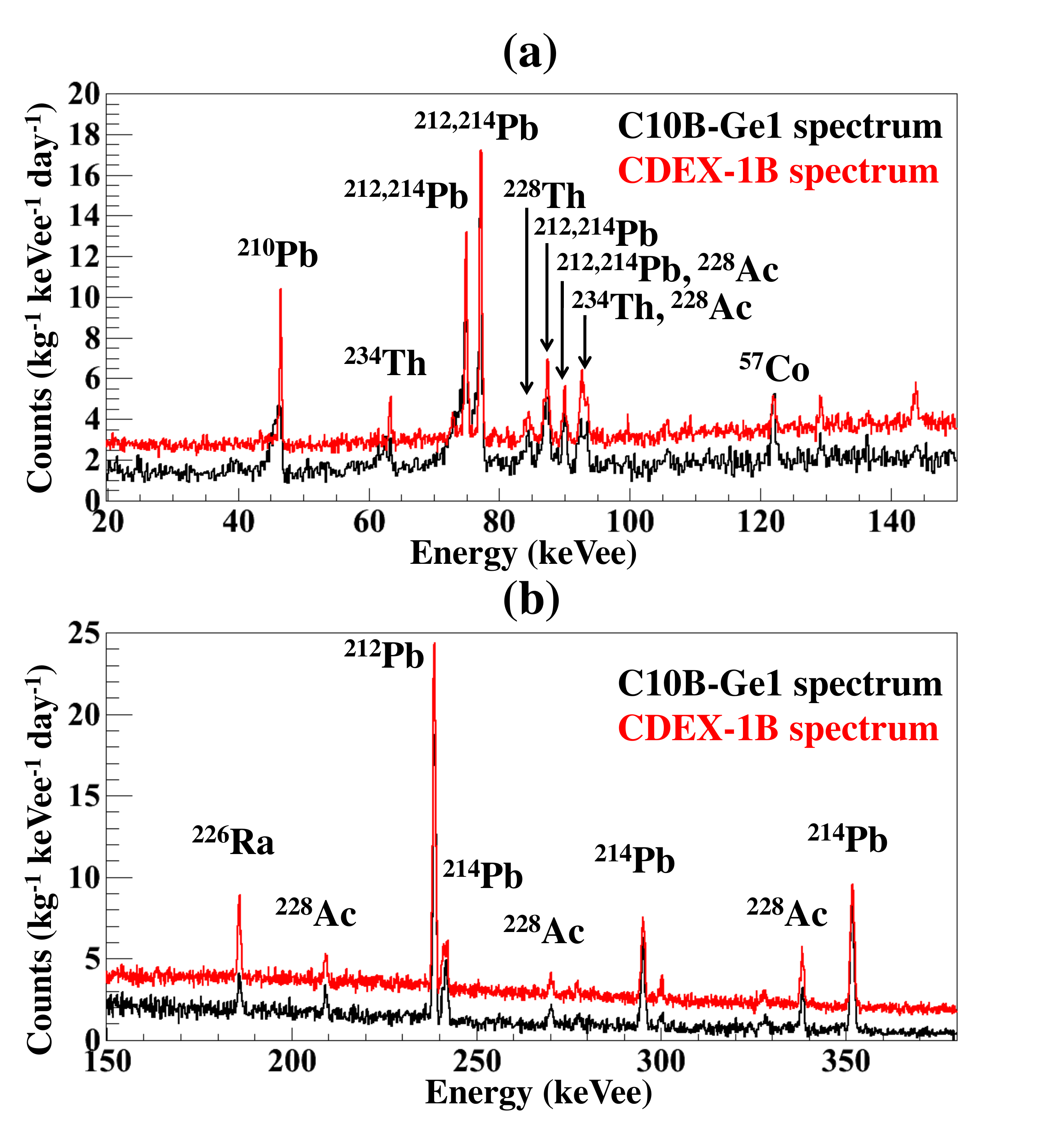}
\caption{Energy spectra of C10B-Ge1 in the 20--150 keVee and 150--350 keVee ranges. The energy spectra of CDEX-1B without an anti-Compton cut are shown for comparison.}
\label{fig::spectrum_high}
\end{figure}

The following peaks (listed in Table~\ref{tab:highpeaks}) were identified in the high-energy region: (1) isotopes from the $^{238}$U and $^{232}$Th decay chains and (2) internal cosmogenic $^{57}$Co. The background spectra of the C10B-Ge1 detector in the energy ranges 20--150 keVee and 150--350 keVee are shown in Figure~\ref{fig::spectrum_high}. The energy spectra of CDEX-1B without an anti-Compton cut are shown for comparison.

\begin{table}[!htbp]
\begin{ruledtabular}
\caption{Identified KX and $\gamma$-rays in the background spectrum of the C10B-Ge1 detector in the high-energy region.}
\label{tab:highpeaks}
\centering
\begin{tabular}{lccc}
Energy (keV) & Types of rays & Isotope & Source \\
\hline
46.5  & $\gamma$ & $^{210}$Pb &  $^{238}$U \\
63.3  & $\gamma$ & $^{234}$Th &  $^{238}$U \\
74.8  & $k\alpha2$ & $^{212,214}$Pb &  $^{238}$U,$^{232}$Th \\
77.1  & $k\alpha1$ & $^{212,214}$Pb &  $^{238}$U,$^{232}$Th \\
84.4  & $\gamma$ & $^{228}$Th &  $^{232}$Th \\
86.8  & $k\beta3$ & $^{212,214}$Pb &  $^{238}$U,$^{232}$Th \\
87.3  & $k\beta1$ & $^{212,214}$Pb &  $^{238}$U,$^{232}$Th \\
89.8  & $k\beta2$ & $^{212,214}$Pb &  $^{238}$U,$^{232}$Th \\
90.0  & $k\alpha2$ & $^{228}$Ac &  $^{232}$Th \\
92.4  & $\gamma$ & $^{234}$Th &  $^{238}$U \\
92.8  & $\gamma$ & $^{234}$Th &  $^{238}$U \\
93.4  & $k\alpha1$ & $^{228}$Ac &  $^{232}$Th \\
122.1  & $\gamma$ & $^{57}$Co & cosmogenic isotope  \\
186.2  & $\gamma$ & $^{226}$Ra &  $^{238}$U \\
209.3  & $\gamma$ & $^{228}$Ac &  $^{232}$Th \\
238.6  & $\gamma$ & $^{212}$Pb &  $^{232}$Th \\
242.0  & $\gamma$ & $^{214}$Pb &  $^{238}$U \\
270.2  & $\gamma$ & $^{228}$Ac &  $^{232}$Th \\
295.2  & $\gamma$ & $^{214}$Pb &  $^{238}$U \\
338.3  & $\gamma$ & $^{228}$Ac &  $^{232}$Th \\
351.9  & $\gamma$ & $^{214}$Pb &  $^{238}$U \\
\end{tabular}
\end{ruledtabular}
\end{table}

\section{Bulk/surface event discrimination and extremely-fast events}

Being a $p$PCGe detector, C10B-Ge1 has an $n^+$ surface layer with a thickness of (0.88$\pm$0.12) mm~\cite{deadlayer}. Meanwhile, the SiO$_x$ passivated layer at the surface of the $p^+$ point-contact electrode is only sub-hundred nm thick. Events depositing energy in the \emph{n$^{+}$} surface layer, namely, surface events, will generate a slow rising pulse and an incomplete charge collection, owing to the weak electric field and severe recombination of electron--hole pairs in this region~\cite{LiHB_2014a,gerda1,majorana}. Consequently, only bulk events are retained in the physical analysis. 

\begin{figure}[!htbp]
\centering\includegraphics[width=0.95\columnwidth]{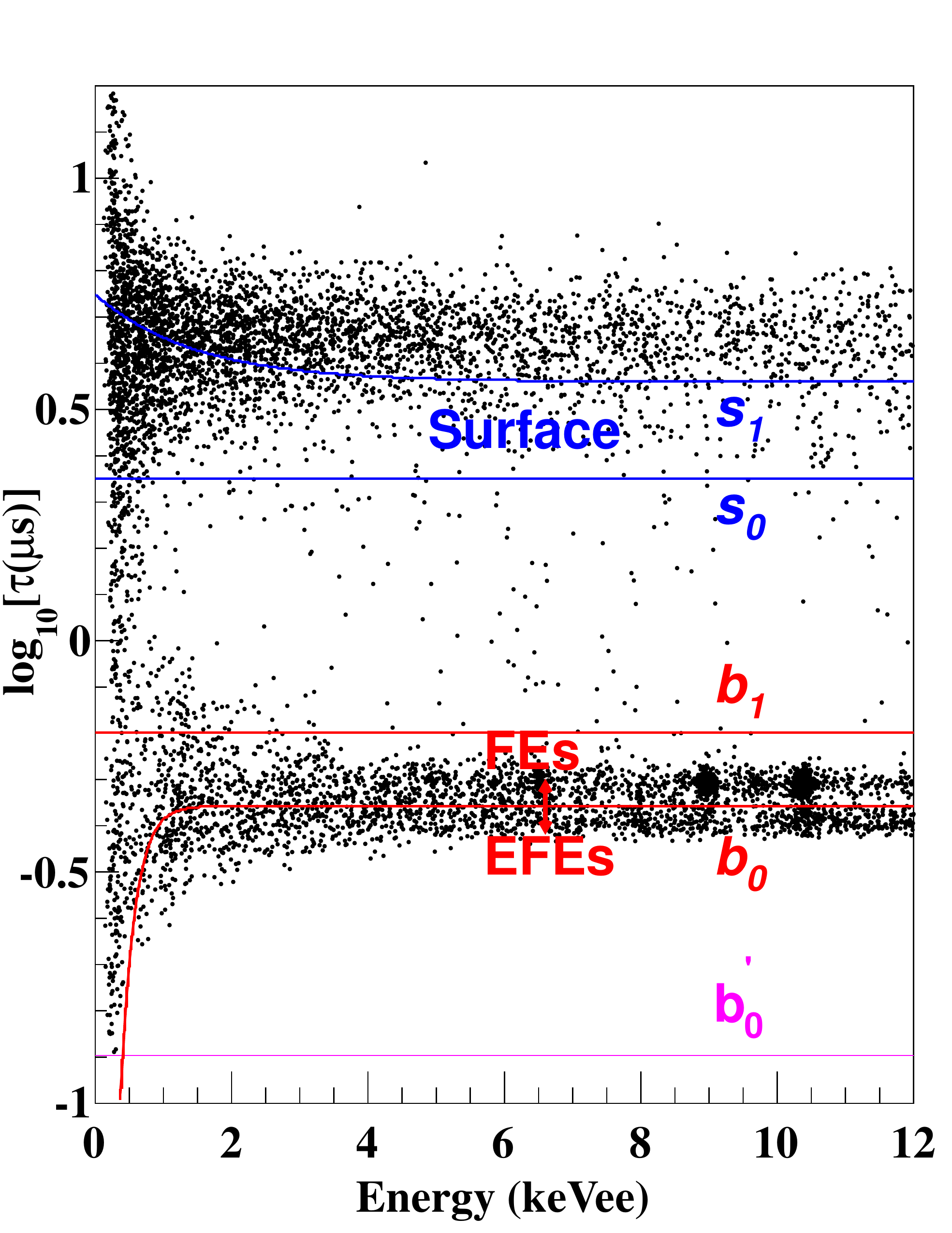}
\caption{Scatter plot of the rise-time (log$_{10}(\tau$)) vs. the measured energy of dark-matter search events. The input parameters were the four boundaries (b$_0$, b$_1$, s$_0$, s$_1$) related to the approximately pure bulk and surface regions. The red line b$_0$ was obtained by fitting the best normalization interval, and the purple line b$_{0}^{'}$ is a test curve to evaluate the b$_0$ selection.}
\label{fig::risetime_energy}
\end{figure}

Bulk and surface events are discriminated by the rise-time $\tau$, which is measured by fitting the $T_{p}$ pulse to a hyperbolic tangent function~\cite{LiHB_2014a,cdex12014,cdex12016,cdex12018}. Figure~\ref{fig::risetime_energy} plots the log$_{10}(\tau$) distribution vs. the measured energy of {\it in situ} events. The two-band structure of bulk and surface events is well separated above 1.5 keVee. However, at lower energies, the bulk and surface events are merged by the electronic-noise smearing effect. To determine the actual number of bulk events ($B_r$), we developed the Ratio Method~\cite{ratiomethod,cdex12018}, which assumes that the dark-matter search data and calibration source data share common probability density functions (PDFs) of the bulk/surface rise-time distribution. When reconstructing $B_r$, the important input parameters are the four boundaries ($b_0$, $b_1$, $s_0$, $s_1$) related to the approximately pure bulk and surface region. In the CDEX-1B analysis, these four boundaries were set as four energy-independent constants ~\cite{cdex12018}.

As the rise-time resolution of the C10B-Ge1 detector is superior to that of CDEX-1B, a category of extremely-fast events (EFEs) with faster rise-time in the lower part of the bulk band appears in Figure~\ref{fig::risetime_energy}. Thus, the C10B-Ge1 detector identifies two categories of bulk events: fast events (FEs) and EFEs. 

\begin{figure}[!htbp]
\centering\includegraphics[width=\columnwidth]{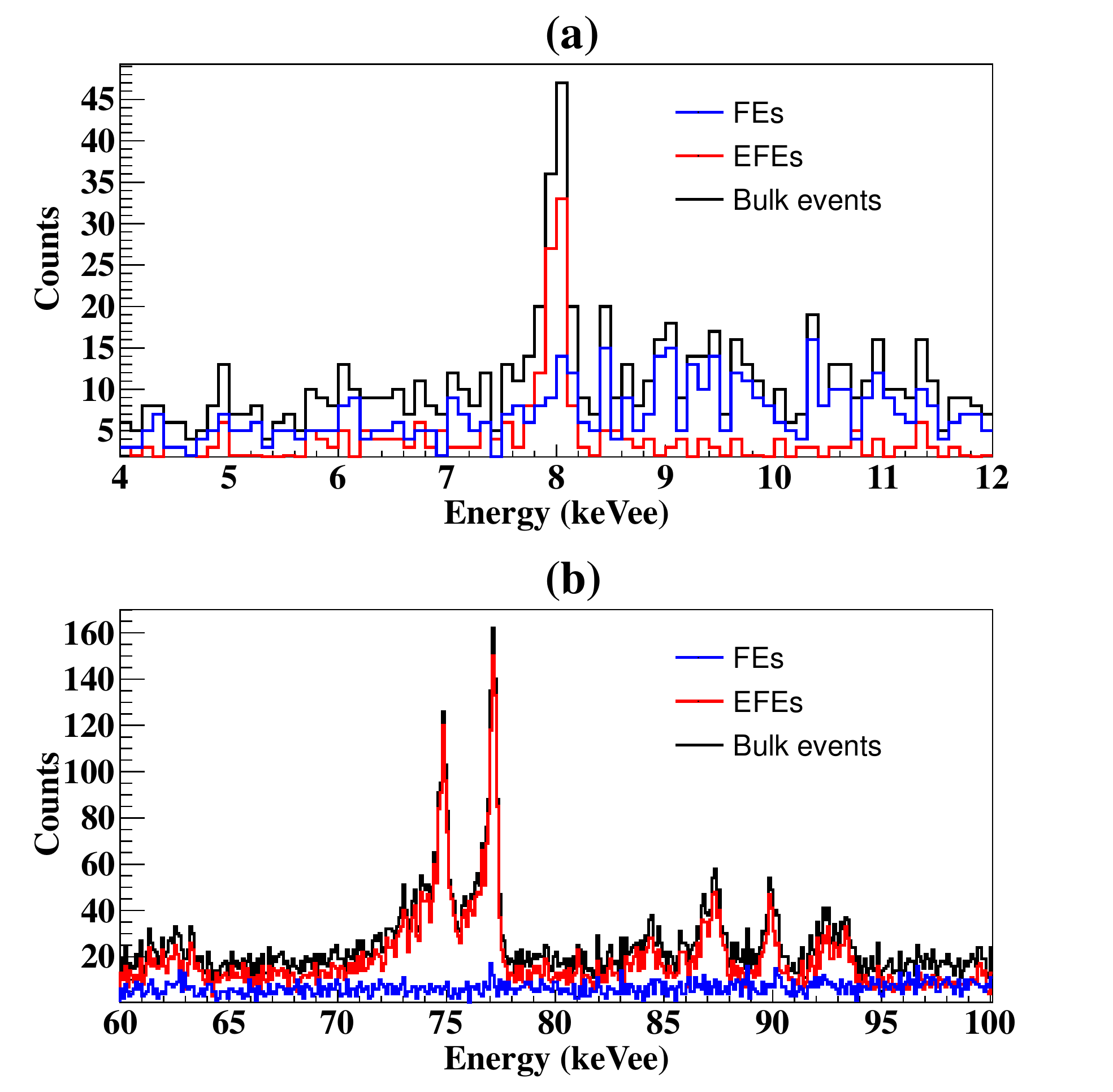}
\caption{Bulk event, Fe, and EFE spectra obtained from (a) $^{109}$Cd calibration data (4--12 keVee) and (b) dark-matter search data (60--100 keVee).}
\label{fig::vbulk}
\end{figure}

These EFEs are actually expected~\cite{gerda1}, and have been verified via electric field calculations of germanium crystals combined with Monte Carlo simulations. They mainly arise in the vicinity of the $p^+$ point electrode of the $p$PCGe detector. According to the analysis of the background spectrum (Figure~\ref{fig::spectrum_low}), the X-rays at 8.0 keV and 8.6 keV originate from $\gamma$-ray excitation of outside materials, Cu and Zn. To further confirm the origin of EFEs, we analyzed the experimental spectrum of the C10B-Ge1 detector irradiated by a $^{109}$Cd source. The EFEs were selected by a simple rise-time cut of $\tau<$0.44~$\mu$s at $>$4 keVee. Figure~\ref{fig::vbulk}(a) depicts the spectra of the bulk events containing FEs and EFEs. The Cu X-rays appear almost exclusively in the EFE spectrum. Figure~\ref{fig::vbulk}(b) depicts the background spectra in the high-energy region. X-rays (70--95 keVee) from $^{212,~214}$Pb, $^{228,~234}$Th, and $^{228}$Ac appear mainly in the EFE band. These two phenomena are consistent with the abovementioned verification, indicating that a considerable number of low-energy events originate from the excitation of outside materials, leading to energy deposition close to the $p^+$ electrode. By identifying these dark-matter search events, we can better understand the background sources and further increase the detection sensitivity of DM.

The $p^+$ electrode surface lacks a dead layer that would absorb the radiation originating from the electronics, shielding and structural materials near the point-electrode area, and the contamination due to the alpha emitters of the $p^{+}$ surface itself. The ratio of EFEs to the total number of dark-matter search events is larger at low energies as compared with higher energies. However, the $\gamma$-rays of the calibration sources located at the side or on top of the detector are impeded by the cryostat housing; hence, they are less likely to reach the $p^+$ electrode vicinity. Consequently, the number of EFEs is lower than that in the dark-matter search data. 

Considering the different ratios of FEs and EFEs in the dark-matter search and calibration data, the EFEs were first removed by an EFE cut (log$_{10}(\tau)$$<$$b_0$). The remaining data satisfied the common bulk PDF assumption of the Ratio Method. The EFEs were returned to the bulk counts following the bulk/surface correction procedure.

As shown in Figure~\ref{fig::risetime_energy}, the FEs and EFEs are well separated at energies above 4 keVee, but are difficult to distinguish below 4 keVee. At lower energies, they are smeared by electronic noise. In this analysis, the EFE cut line $b_0$ in Figure~\ref{fig::risetime_energy} was fitted as an exponential function of energy. The curve was derived by fitting the best normalization interval of each 500-eVee energy bin, starting at 160 eVee. Here, the statistics were rendered as significant as possible while maintaining consistent rise-time distributions of the events. The curve contains all of the FEs and EFEs at energies below 0.5 keVee because the two event categories cannot be differentiated in this energy region.

\begin{figure}[!htbp]
\centering\includegraphics[width=\columnwidth]{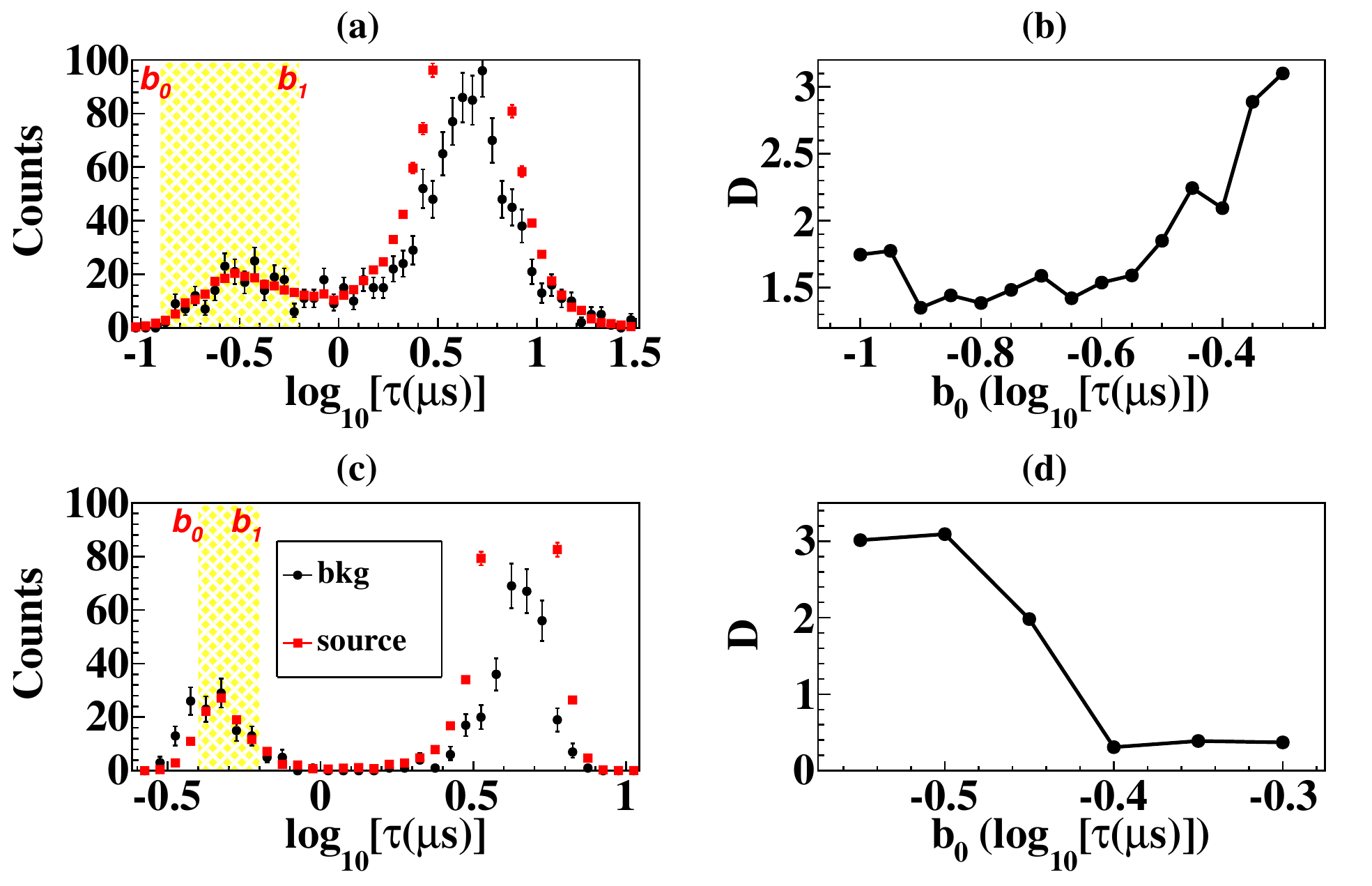}
\caption{Normalization interval and scan results of b$_{0}$ in the 0.16--0.66 keVee energy bin (a, b) and the 1.66--2.16 keVee energy bin (c, d).}
\label{fig::normalization_interval}
\end{figure}

The best normalization interval was determined by the following steps:

(1) Within a certain energy bin, e.g., 0.16--0.66 keVee and 1.66--2.16 keVee in Figure~\ref{fig::normalization_interval}(a) and Figure~\ref{fig::normalization_interval}(c), respectively, calculate the deviation of each rise-time bin in the normalization interval between the source and dark-matter search data. The formula is given by
\begin{eqnarray}
\overline{N_{j}}=\sum_{i=1}^{2}\frac{N_{i,j}}{\sigma_{i,j}^{2}}/\sum_{i=1}^{2}\frac{1}{\sigma_{i,j}^{2}},\nonumber\\
A_{j}=\sum_{i=1}^{2}\frac{1}{\sigma_{i,j}^{2}}(N_{i,j}-\overline{N_{j}})^{2},
\label{eq:normalization interval}
\end{eqnarray}
where $N_{i,j}$ represents the count rate of the dark-matter search data ($i$=1) or the source data ($i$=2) in the $j$-th rise-time bin, $A_{j}$ represents the deviation of the $j$-th rise-time bin, and $\sigma_{i, j}$ is the statistical error.

(2) Calculate the deviation over the whole normalization interval as follows:
\begin{eqnarray}
D=\frac{1}{n}\sum_{j=1}^{n}A_{j},
\label{eq:mean deviation}
\end{eqnarray}
where $D$ represents the deviation of the whole normalization interval, and $n$ represents the number of rise-time bins in that normalization interval.

(3) Find the best normalization interval by scanning b$_0$ in every 500-eVee energy bin. The principle is to maintain (as far as possible) both significant statistics and consistent rise-time distributions of the events. The smaller the $D$ in Eq.~(\ref{eq:mean deviation}), the more consistent are the dark-matter search and source data. In Panels (b) and (d) of Figure~\ref{fig::normalization_interval}, $b_0$ was selected as --0.9 and --0.4, respectively.

\begin{figure}[!htbp]
\centering\includegraphics[width=\columnwidth]{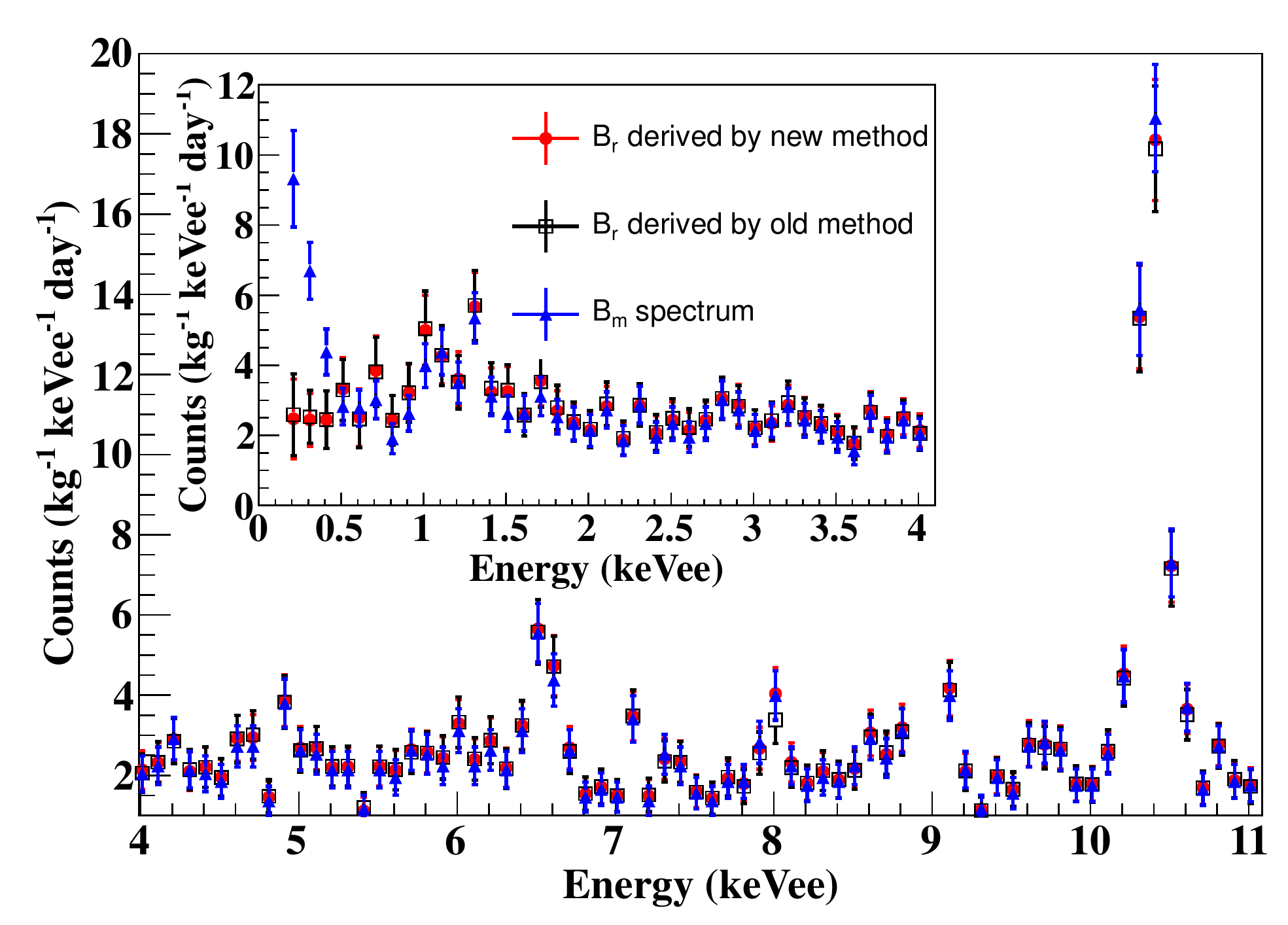}
\caption{Comparison of reconstructed $B_r$ values of the dark-matter search data from C10B-Ge1 using the old and new $b_0$ selection methods. The $B_m$ spectrum is also shown.}
\label{fig::br_spectrum}
\end{figure}

(4) Derive the curve by fitting the best normalization interval in each 500-eVee energy bin, beginning at 160 eVee as shown in Figure~\ref{fig::risetime_energy}.

Figure~\ref{fig::br_spectrum} shows the reconstructed $B_r$ of the dark-matter search data from C10B-Ge1. The effect of this $b_0$ selection (new method) was evaluated in a comparison test against another $b_{0}^{'}$ selection (old method used in CDEX-1B, which assumes a constant value and is plotted as the purple line in Figure~\ref{fig::risetime_energy})~\cite{cdex12018}. The result is shown in Figure~\ref{fig::br_spectrum}. The two spectra differ slightly at the energy peak of the Cu X-rays (8.0 keV), wherein most events are EFEs.

\section{Summary}
The second-generation CDEX experiment, CDEX-10, has successfully searched light DM using germanium detectors immersed in LN$_{2}$. A DAQ system was set up to digitize and record various original detector waveforms. The C10B-Ge1 detector achieved a series of excellent performances: an ultra-low physics analysis threshold of 160 eVee, a low background level of 2.5 cpkkd in the 1.5--2.0 keVee energy range, and high energy resolution (219 eVee FWHM at 10.37 keV).

By virtue of the better rise-time resolution in C10B-Ge1 as compared to that in CDEX-1A and CDEX-1B~\cite{cdex12016,cdex12018}, we also clarified a category of EFEs with extremely fast rise-times. The EFEs were experimentally confirmed to originate from X-rays penetrating the $p^+$ passivated layer. To handle the EFEs, the Ratio Method was improved~\cite{cdex12018} by removing the EFEs prior to bulk/surface discrimination, and reinstating them to determine the $B_r$ spectrum.

With such excellent performance, the first physical run of the CDEX-10 experiment extended from February 26, 2017 to November 7, 2017, and the first results from the 102.8-kg day dark-matter search dataset of the C10B-Ge1 detector have been distributed. Thus far, the CDEX collaboration has achieved the highest physical sensitivity for $\chi$-nucleon spin-independent interactions in the 4--5 GeV energy region~\cite{cdex10-PRL}. 

\begin{acknowledgments}
This work was supported by the National Key Research and Development Program of China (No. 2017YFA0402201) and the National Natural Science Foundation of China (No.11475092, 11475099, 11675088, 11725522).
\end{acknowledgments}

\end{document}